\documentclass[10pt,letterpaper]{article}
\usepackage{opex3}
\usepackage{color}
\usepackage{threeparttable}
\usepackage[utf8]{inputenc}
\usepackage[english]{babel}
\usepackage{amsmath}
\usepackage{amssymb}
\usepackage{braket}
\usepackage{cite}

\addto\captionsenglish{}

\begin{document}
\renewcommand{\refname}{References and links}

\title{On-chip generation of photon-triplet states}

\author{Stephan Krapick,$^{1,2}$ Benjamin Brecht,$^{1}$ Harald Herrmann,$^{1}$ Viktor Quiring,$^{1}$ and Christine Silberhorn$^{1,3}$}

\address{$^1$Applied Physics, University of Paderborn, Warburger Strasse 100, 33098 Paderborn, Germany\\
$^2$stephan.krapick@gmail.com\\
$^3$christine.silberhorn@uni-paderborn.de}

\begin{abstract}
Efficient sources of many-partite non-classical states are key for the advancement of quantum technologies and for the fundamental testing of quantum mechanics. We demonstrate the generation of time-correlated photon triplets at telecom wavelengths via pulsed cascaded parametric down-conversion in a monolithically integrated source. By detecting the generated states with success probabilities of $(6.25\pm1.09)\times10^{-11}$ per pump pulse at injected powers as low as $10\;\mu\mathrm{W}$, we benchmark the efficiency of the complete system and deduce its high potential for scalability. Our source is unprecedentedly long-term stable, it overcomes interface losses intrinsically due to its monolithic architecture, and the photon-triplet states dominate uncorrelated noise significantly. These results mark crucial progress towards the proliferation of robust, scalable, synchronized and miniaturized quantum technology.
\end{abstract}


\section{Introduction}

Photons serve as excellent information carriers due to their low decoherence and weak interactions with matter. The creation of high-dimensional (entangled) photon states, such as tripartite Greenberger-Horne-Zeilinger-states (GHZ) \cite{Greenberger1990}, is desirable for proving deterministically the non-classical nature of quantum physics as a complete theory \cite{Einstein1935}, but it requires sophisticated quantum technologies to do so \cite{Shalm2015,Giustina2015}. 

Many recent developments in quantum optics build on the benefits of robust and compact integrated circuits as parts of complex quantum networks. Integrated devices with multiple functionalities have been successfully demonstrated, e. g., in the fields of photon entanglement \cite{Politi2008,Martin2012a,Jin2014}, quantum interference \cite{Silverstone2014} and boson sampling \cite{Crespi2013,Spagnolo2014}. The technological challenge to combine multiple functionalities in a mutually compatible manner remains. 

To date, photonic tripartite states have been generated successfully via simultaneous \cite{Keller1998,Barz2010,Guerreiro2014} and cascaded parametric down-conversion (PDC) \cite{Huebel2010,Shalm2013,Hamel2014}, by cascading three- and/or four-wave mixing processes \cite{Jia2012,Qin2014,Jing2014,Ding2015} or by using tri-exciton decays in coupled solid-state quantum dot sources \cite{Khoshnegar2015}. Also, the generation of photon triplets using cascaded superlattices in nonlinear crystals has been studied in detail in \cite{Antonosyan2011}. Due to the chosen architectures, most of the experimental approaches inherently suffer from loss at the interfaces of the involved elements, they are space-consuming and can be susceptible to long-term stability issues. Additionally, the required pump powers for photon-triplet generation are typically of the order of several milliwatts, whereas more energy-efficient systems are favorable for real-world applications.

Based on the idea of cascaded PDC \cite{Huebel2010}, we pursue a fully monolithic approach to generate photon-triplet states on a second-order nonlinear waveguide chip. We use lithium niobate with diffused waveguide structures \cite{Schmidt1974,Parameswaran2002}, since they offer low loss \cite{Regener1985}, high source brightness \cite{Tanzilli2001,Krapick2013,Krapick2014} and fast electro-optical switching capabilities \cite{Jin2014} for reconfigurable quantum optical applications. By contrast to schemes that utilize continuous-wave pump lasers, we deploy pump pulses, which makes our source compatible to synchronized quantum network architectures.

\section{Device design and theoretical implications}
\label{sec:device}

\begin{figure}[t]
\centering
\includegraphics[width=0.95\linewidth]{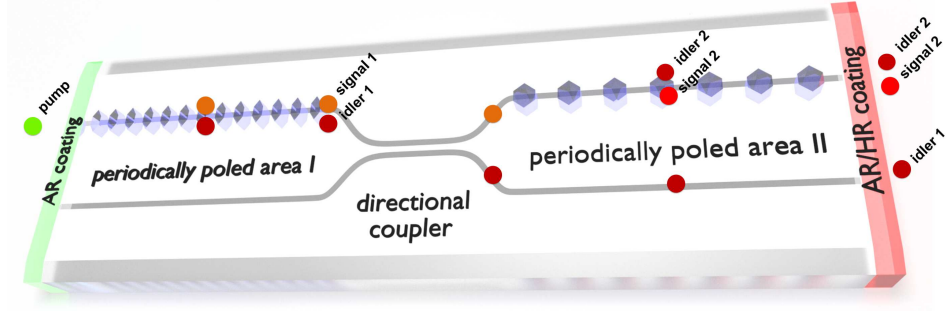}
\caption{Device layout and on-chip functionalities: picosecond pulses at $532\,\mathrm{nm}$ are injected to the coupled waveguide structure and pump the cascaded PDC process. They decay to photons at $790.3\,\mathrm{nm}$ (signal 1) and at $1625\,\mathrm{nm}$ (idler 1) in the periodically poled area I. The integrated directional coupler splits up the generated photon pairs spatio-spectrally. While idler 1 photons couple to the adjacent waveguide, the corresponding signal photons remain in the original waveguide and decay to secondary photon pairs at $\left(1551\pm25\right)\,\mathrm{nm}$ (signal 2) and $\left(1611\pm25\right)\,\mathrm{nm}$ (idler 2) in the periodically poled area II. Anti-reflective dielectric coatings on the waveguide end-faces provide the reduction of Fresnel-losses at telecom wavelengths, while the green pump is reflected in order to reduce additional filtering efforts.}
\label{triplet-chip}
\end{figure}

We designed and fabricated an integrated device, which is illustrated and explained in Fig. \ref{triplet-chip}. A coupled-waveguide structure of constant waveguide width and based on titanium-diffusion \cite{Schmidt1974,Krapick2013} has been introduced to a $76\,\mathrm{mm}$ long lithium niobate chip. In front of and behind the integrated wavelength division multiplexer (WDM) we have implemented two differently poled structures, which act as guided-wave PDC sources, by periodically inverting the nonlinear susceptibility using a pulsed electric-field-poling technique. Although our waveguides in principle support both polarizations, we restrict ourselves here to PDC processes, where only TM modes are involved, since the highest nonlinear coefficient $d_{33}$ can be deployed for both down-conversions. The choice of only one polarization also implies that the designed integrated WDM has to act only as a wavelength demultiplexer, but not as a polarization-wavelength-splitting element.

The actual photon-triplet generation process is considered as follows: picosecond pulses at $\lambda_\mathrm{p}=532\,\mathrm{nm}$ serve as the pump and drive the first type-0 PDC process in order to produce pairs comprising signal 1 photons (s1) at $\lambda_\mathrm{s1}=\lambda_\mathrm{p2}=790.3\,\mathrm{nm}$ and idler 1 photons (i1) at $\lambda_\mathrm{i1}=1625\,\mathrm{nm}$. The WDM separates the generated primary photons pairs with high probability in a spatio-spectral manner. The signal 1 photons remain in the original arm, whereas the idler 1 photons are transferred to the adjacent waveguide. By deploying the signal 1 photons as the pump (p2) in the second type-0 PDC stage we generate secondary signal photons (s2) at $\lambda_\mathrm{s2}=\left(1551\pm25\right)\,\mathrm{nm}$ and idler photons (i2) at $\lambda_\mathrm{i2}=\left(1611\pm25\right)\,\mathrm{nm}$. In summa, we are able to create three time-correlated photons in the telecom wavelength regime with the cascaded PDC processes, while the energy conservation, $\hbar\omega_\mathrm{p}=\hbar\omega_\mathrm{i1}+\hbar\omega_\mathrm{s2}+\hbar\omega_\mathrm{i2}$, must be fulfilled. Likewise, the wavelength-dependent conservations of momenta in both PDC processes (commonly called phase-matchings) have to be made mutually compatible as described in Appendix A. Note that our compact monolithic approach provides intrinsic spatial mode matching and conveniently tunable spectral mode matching of the intermediate signal 1/pump 2 photons along the coupled-waveguide structure. Additionally, our titanium-diffused waveguides exhibit very low propagation loss of $\sim0.08\,\mathrm{dB/cm}$ on average at telecom wavelengths.

The first PDC stage in the cascade is pumped with a classical field. It is known that parametric down-conversion processes, which are not pumped by single-photons, produce not only single, but also higher-order photon pairs \cite{Takesue2010}. Thus, we must expect to generate a statistical mixture of genuine photon triplets, $\ket{\psi_\mathrm{triplet}}=\ket{1\,1\,1}$, and states including higher-order photon contributions in our process. This means that we also generate states of the form $\ket{\psi_\mathrm{m-plet}}=\ket{m\,1\,1}$ with certain probabilities. The mean photon number per optical pulse behind the primary PDC is denoted by $\langle m\rangle$. We write for the vector containing the photon-number-occupation probabilities of the first PDC process:
\begin{equation}
\label{eq:01}
\boldsymbol{\rho}=
 \begin{pmatrix}
 \rho_0\\
 \vdots\\
 \rho_n
 \end{pmatrix},
\end{equation}
where the vector components $\rho_m\in\{\rho_0,...,\rho_n\}$ are the probabilities to generate $m$ PDC photon pairs, and $0\le m\le n$ is an integer number. For our case of spectrally multi-mode PDC, the photon probabilities obey Poisson statistics and are, thus, given by
\begin{equation}
\label{eq:02}
\rho_m=\frac{e^{-\langle m \rangle}\langle m\rangle^{m}}{m!}, \;m\in\mathbf{N_0}.
\end{equation}
Because the mean photon pair number per optical pulse can be written as
\begin{equation}
\label{eq:03}
\langle m\rangle=\sum\limits_{m=0}^\infty m\rho_m,
\end{equation}
a reasonable photon-triplet generation approach must provide $\langle m\rangle\ll1$ for the primary PDC process. Given that case, the higher-order photon contributions are significantly reduced such that $\sum_{m\ge2}^\infty m\rho_m\ll\rho_1$, meaning that we pump the secondary PDC stage almost exclusively with single photons. Thus, at low pump powers, we will measure mainly genuine photon triplets in the overall process. Reference \cite{Krapick2015} provides an in-depth theoretical analysis of quantitative measures for state preparation fidelities based on observed experimental parameters.

\section{Long-term stable experimental setup}

\begin{figure}[t]
\centering
\includegraphics[width=0.95\linewidth]{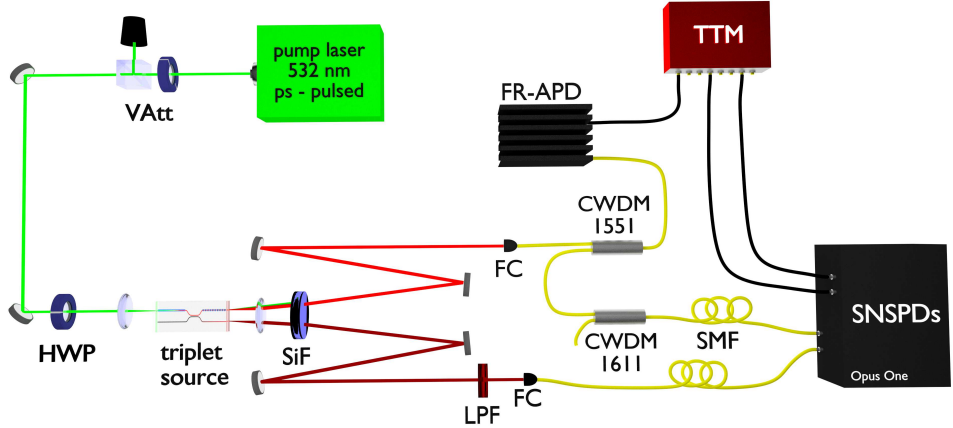}
\caption{Experimental setup for photon-triplet verification: the tripartite states reside in the two output beams behind the device. A silicon filter blocks the pump and photons at $790.3\,\mathrm{nm}$ in both beams. The lower output comprises the primary idler photons and the upper beam contains secondary photon pairs. We separate the secondary photon pairs quasi-deterministically according to their spectral correlations by fiber-based coarse wavelength division multiplexers at $\left(1551\pm7\right)\,\mathrm{nm}$ and $\left(1611\pm7\right)\,\mathrm{nm}$, respectively. The three resulting fiber-coupled beams address free-running binary detectors. The data acquisition and evaluation is done by a time-tagging module and appropriate software. Legend: CWDM - coarse wavelength division multiplexer, FC - fiber coupling stage, FR-APD - free-running avalanche photodiode, HWP - half-wave plate, LPF - long-pass filter, SiF - silicon filter, SMF - single-mode fiber, SNSPD - superconducting nanowire single photon detector, TTM - time-tagging module, VAtt - variable attenuator.}
\label{triplet-setup}
\end{figure}

For testing our device, we implement the setup shown in Fig. \ref{triplet-setup}. Thermal stabilization of our integrated chip at temperatures of $\theta=164.8^\circ\mathrm{C}$ ensures that the intermediate signal 1 wavelength is $\lambda_\mathrm{s1}=\lambda_\mathrm{p2}=790.3\,\mathrm{nm}$, which is required for non-degenerate secondary PDC generation (see Appendix A for detailed explanation). This setting allows for optimum mutual compatibility of the two PDC processes, and it improves the spatio-spectral separability of secondary photon pairs by using fiber-based coarse wavelength division multiplexers (CWDM). These standard components also serve as high-performance filters for noise events as well as for parasitic photons from the primary PDC stage. Otherwise, those photons could affect the detection of secondary PDC photons, because the first PDC process happens around ten million times more often than the second one.

\begin{figure}[t]
\centering
\includegraphics[width=0.75\linewidth]{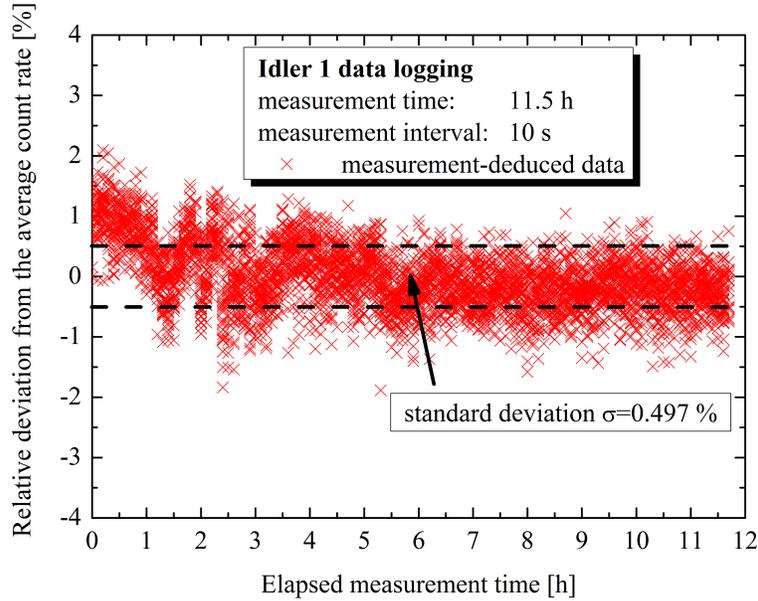}
\caption{Long-term stability of our device: we recorded the idler 1 count rate (per second) every ten seconds over the whole measurement time of 11.5 hours. The plot shows the relative change of the count rate with respect to the averaged value. We see that the relative fluctuations are less than 2\% over the full measurement duration. After 2.5 hours of elapsed time, there is practically no residual drift, and the individual data points scatter with standard deviation of less than 0.5\% around zero. This result is evidence for the unprecedented long-term stability of our integrated quantum circuit.}
\label{stability}
\end{figure}

We measure the photon-triplet events using two superconducting nanowire single photon detectors (SNSPDs) with $\eta_\mathrm{det,i1}=0.6$ and $\eta_\mathrm{det,i2}=0.7$ of detection efficiency (Opus One, Quantum Opus/PicoQuant Photonics North America Inc.), as well as one InGaAs avalanche photo diode with $\eta_\mathrm{det,s1}=0.25$ (ID230-SMF-FR, ID Quantique SA) in conjunction with a time-tagging module. Optical and electronic path differences between the three detectors have been compensated for, such that the expected, time-correlated detection events can be registered at around zero delay with respect to each other. The measurements have been performed for $11.5$ hours with very high stability. This is indicated by the plot in Fig. \ref{stability}, where we show the relative change of the idler 1 count rate with respect to the average value. We chose the idler 1 detection events for monitoring the stability, because they occur orders of magnitudes more often and with significantly better signal-to-noise ratios than secondary PDC detection events. The single event rates of signal 2 and idler 2 are orders of magnitudes lower than the respective detector noise count rates and have not been considered as stability indicators.

The continuous-wave-equivalent pump power of $P_\mathrm{pump}=\left(10.0\pm0.1\right)\,\mu\mathrm{W}$ at a repetition rate of $10\,\mathrm{MHz}$ corresponds to a mean photon pair number per pulse of only $\langle m\rangle=0.215\pm0.02$ behind the first PDC stage. Thus, we expect that predominantly $\left(\sim88\,\%\right)$ genuine photon-triplet states $\ket{\psi_\mathrm{triplet}}=\ket{1,1,1}$ are generated. Likewise, we deduce that a high conversion efficiency in our primary PDC stage limits the cleanliness of the generated photon-triplets by generating higher-order photon pairs \cite{Krapick2015}.

We chose the signal 2 photon detection as the reference events for our data analysis, because these occur at the lowest detection rates due to the InGaAs detector efficiency. A three-fold coincidence is given, if a signal 2 detection event announces the detection of its idler 2 twin photon, and if the corresponding idler 1 photon is also registered. Hence, the relative arrival times for idler 1 events are labeled $\tau_1-\tau_2$, whereas $\tau_3-\tau_2$ denotes the relative arrival time of idler 2 photons with respect to the signal 2 photons. This pseudo-heralding method significantly reduces the computational effort for the post-selection. Additionally, we merge the bins of our time-tagging module sixteen-fold in order to include the joint timing jitter of our apparatus. Thus, the time-bins for the data analysis have widths of $\left(1.317\pm0.002\right)\,\mathrm{ns}$ in both temporal directions.

\section{Results and discussion}

\begin{figure}[t]
\includegraphics[width=0.42\linewidth]{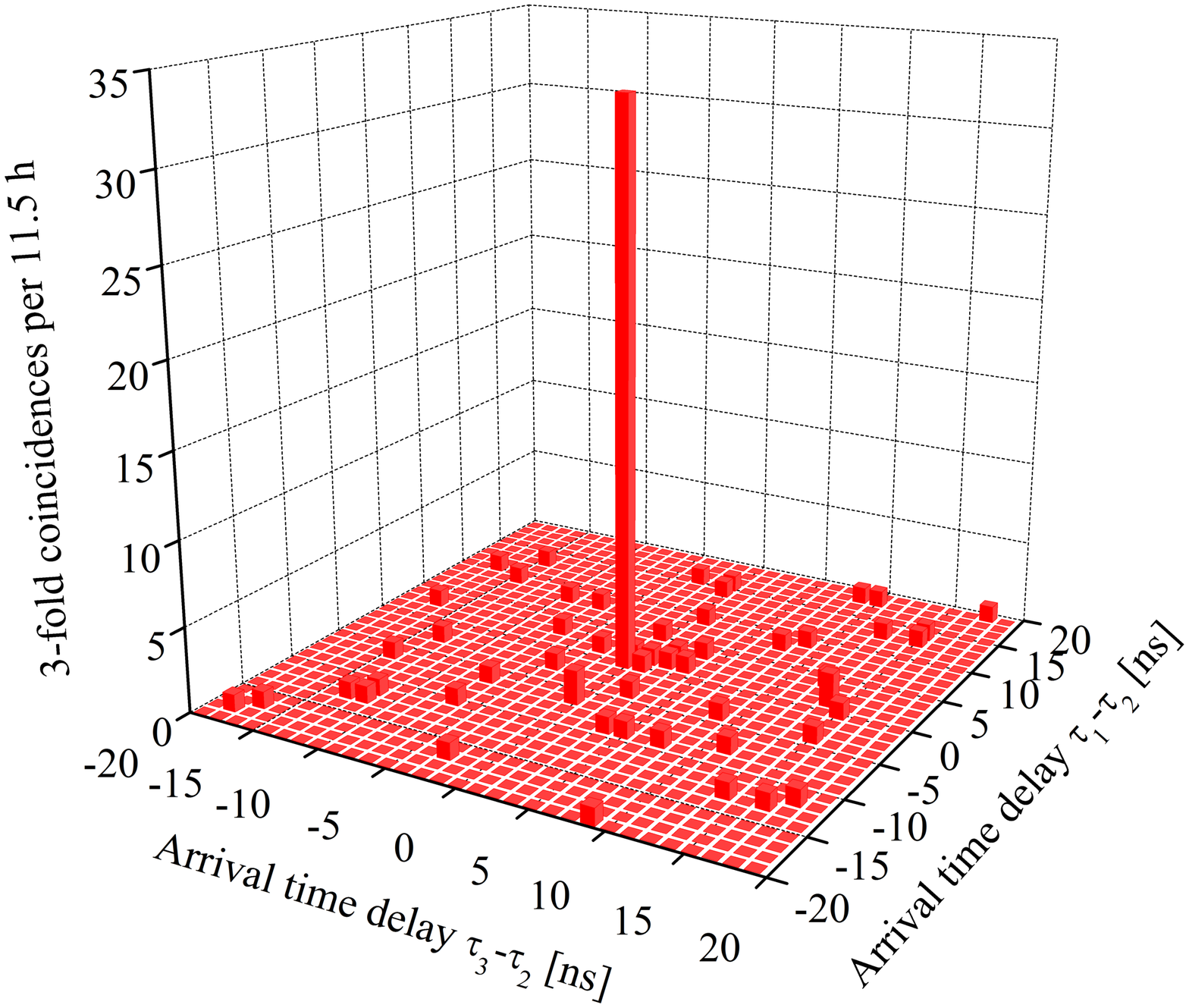}
\includegraphics[width=0.58\linewidth]{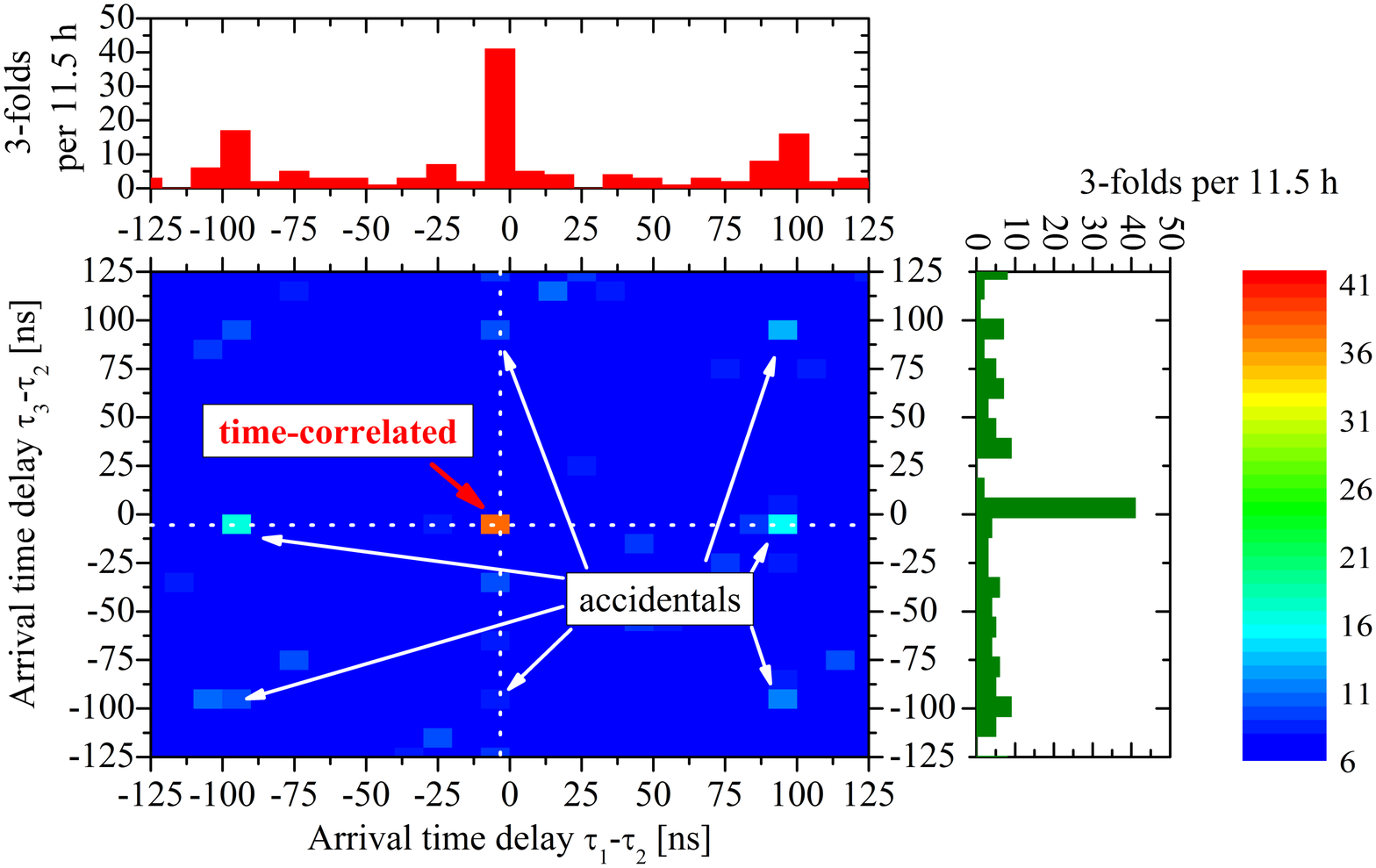}
\caption{Measured three-fold coincidences around the expected arrival times (left). The central peak contains three-fold coincidences, which overcome the average noise background by a manifold and indicate strong time-correlations. We infer raw three-fold coincidence rates of $33$ per 11.5 hours. Note that we merged the acquired data to time bins of $\tau_\mathrm{16bin}=\left(1.317\pm0.002\right)\,\mathrm{ns}$, in order to take the joint timing jitter of the detection apparatus into account. Comparison of the absolute bin occupation in two temporal dimensions for a large analysis window (right). Our pump laser runs at a peak-to-peak repetition time of $100\,\mathrm{ns}$. The expected photon triplets reside in the time bin at around zero time delay between the three detectors. Neighboring three-fold coincidences with significantly lower absolute frequency are also present at timing distances of multiple integers of the inverse laser repetition rate. These accidental triple-coincidences represent the impact of higher-order photons and nonlinear Cherenkov-type PDC on the measurement. Note that we merged our data to time bins of around $10.5\,\mathrm{ns}\times10.5\,\mathrm{ns}$ for improved visualization. The adjacent bar charts stem from the cross-cuts along the white dotted lines in the color-coded graph.}
\label{16-bin}
\end{figure}

Our data analysis benefits from the pulsed pump, which allows us to distinguish between time-correlated photon triplets and noise-related three-fold coincidences. The latter appear due to dark counts of the detector and the blackbody radiation emitted by our heated integrated device. The effect of noise is shown in Fig. \ref{16-bin} (left) for a time window of about $40\,\mathrm{ns}\times40\,\mathrm{ns}$. At relative arrival time delays between individual detection events of $\tau_3-\tau_2=\tau_1-\tau_2=\left(-0.165\pm0.001\right)\,\mathrm{ns}$, we notice time-correlations as an indicator for photon-triplet detection. However, due to the pulsed operation, this result alone does not prove the generation of genuine photon-triplet states. For the verification we have to make sure that the influence of accidental three-fold-coincidences is negligible.

By analogy to conventional pulsed parametric down-conversion \cite{Krapick2013}, we can deduce the impact of accidentals by extracting three-fold coincidences, where neighboring pulses are involved. For their identification we analyze a larger, $600\,\mathrm{ns}\times600\,\mathrm{ns}$-wide, time window. This corresponds to $\sim210000$ bins of $\left(1.317\pm0.002\right)\,\mathrm{ns}$ width, surrounding the signal 2 detection events. It also implies that, due to the repetition time of our pump laser system of $100\,\mathrm{ns}$ and the sixteen-fold merging of the time-bins, we have access to $41$ neighboring pump pulses within the time window for the estimation of higher-order photons and other accidental contributions to the three-fold coincidence rate.

Indeed, we find accidental three-fold coincidences at multiple integers of the pump repetition time for both temporal directions. We illustrate this in Fig. \ref{16-bin} (right), where only a fraction of the analyzed time window is shown for clarity. The graph shows that higher-order photons contribute to the three-fold coincidences along the $(\tau_1-\tau_2)$-axis, indicating an increase on the primary idler detection probability. The appearance of three-fold coincidences in neighboring pulses along the $(\tau_3-\tau_2)$-axis, where only secondary photons should reside, indicates other parasitic influences: either higher-order idler 1 photons survive the demultiplexing on-chip and the subsequent CWDM-filtering, or primary PDC processes involving higher-order mode combinations produce idler 1 photons at secondary PDC photon wavelengths. It is also possible that nonlinear Cherenkov-type PDC \cite{Suhara2003} is generated in the primary PDC stage with a broad idler distribution in the secondary output arms.

We quantify the impact of accidentals by comparing the number of three-fold coincidences in the center spot of the graph in Fig. \ref{16-bin}, where we suspect our photon-triplets to reside, with the average number of accidental three-folds in the $41$ bins, where neighboring pulses are involved. By division of the two results we infer a coincidences-to-accidentals-ratio of $CAR=9.4\pm1.9$. This means that the influence of higher-order photon contributions and other accidentals from the primary PDC stage is not negligible, but very low.
  
Additionally, we perform a statistical analysis of the $210000$, $1.317\mathrm{ns}$-wide, time-bins in order to answer the question: how many three-fold coincidences occur how often. Our intention for this analysis method is to identify noise-related contributions, genuine photon triplets, and also accidental three-fold coincidences. Besides our expectation, that the photon triplets overcome the noise background significantly, the distribution of the accidentals should also deviate from the noise statistics, because those pseudo-time-correlated accidental events are generated by the same pulsed pump that generates the photon triplets.

\begin{figure}[t]
\centering
\includegraphics[width=0.75\linewidth]{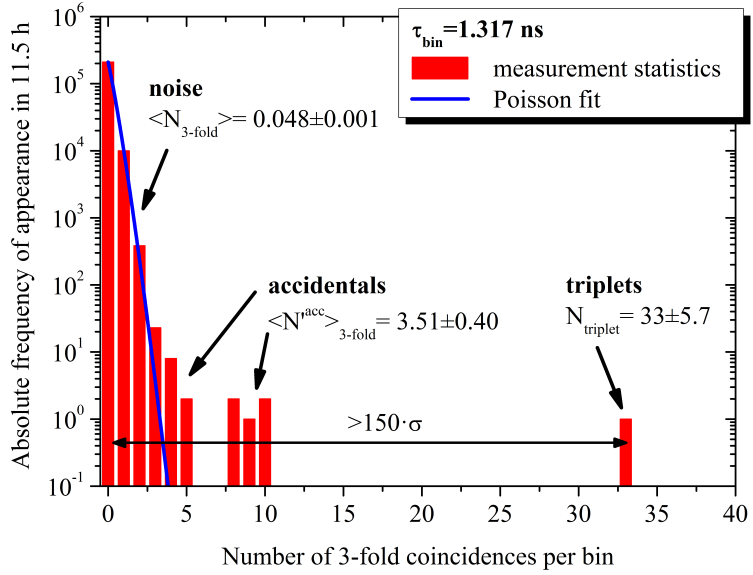}
\caption{Histogram of the absolute frequencies of three-fold coincidences per $1.317\,\mathrm{ns}$-wide time bin: the logarithmic plot exhibits a Poisson-like distribution, underlined by the fit curve. Only one time bin contains $N_\mathrm{triplet}=33\pm5.7$ three-fold coincidences. Compared to the the mean value of $\langle N_\mathrm{3-fold}\rangle=0.048$ (standard deviation $\langle N_\mathrm{3-fold}\rangle^{-1/2}=0.218$), this corresponds to a signal-to-noise-ratio of $SNR>680$. The absolute frequencies for between $4$ and $10$ events per bin indicate pseudo-time-correlated accidental three-fold coincidences.}
\label{statistics}
\end{figure}

In the histogram in Fig. \ref{statistics}, we show the result of our statistical analysis. On the x-axis, we plot the number of three-fold coincidences per time-bin during $11.5$ hours of measurement time. The y-axis shows the absolute frequencies of these events' occurrences. We measured $33$ three-fold coincidences only once with standard deviation of $\sigma_\mathrm{triplet}=5.7$. The noise-related background events are visibly separated and average to $\langle N_\mathrm{3-fold}\rangle=0.048$ three-fold coincidences with a standard deviation of $\sigma_\mathrm{3-fold}=\langle N_\mathrm{3-fold}\rangle^{1/2}=0.218$. The blue line is a Poisson fit of the overall measurement data, the vast majority of which are time bins containing noise. Our result indicates that we are able to detect time-correlated three-fold coincidences with a signal-to-noise-ratio of $SNR>680$. Assuming for now that the measured rate per time-bin of $33$ three-folds stems solely from noise contributions would mean, that it was $150$ standard deviations away from the average noise-related three-fold coincidence rate per bin. In other words: the probability of measuring a noise-related rate of $33$ three-folds per bin in $11.5$ hours is around $p^\mathrm{noise}_\mathrm{3-fold}(33)\approx3.3\times10^{-81}$ and can be considered impossible. Thus, our statistical analysis underlines the strong evidence for time-correlated photon triplets in only one temporal measurement bin. We also notice in the histogram that the accidentals, stemming from neighboring pulses along $\tau_3-\tau_2$ and $\tau_1-\tau_2$, also deviate significantly from the noise-dominated fit curve. This behavior indicates the suspected pseudo-time-correlations of the accidentals due to generation in neighboring pulses. We refer the kind readership to Appendix B, where we provide arguments for the validity of Poisson statistics of noise-related three-fold coincidences.

Our findings verify the generation of $33\pm5.7$ time-correlated photon triplets per $11.5$ hours at the expected relative arrival times. From the absolute number of triplets within the whole measurement duration, we deduce a success probability for the detection of photon triplets $P^\mathrm{exp}_\mathrm{triplet}=\left(6.25\pm1.3\right)\times10^{-11}$ per pump pulse. In order to compare this value with the success probability expected from our experimental circumstances, we take the set pump power and wavelength, $\mathbf{P}_\mathrm{p}$ and $\lambda_\mathrm{p}$, the pump laser repetition rate $R_\mathrm{rep}$ and the efficiencies of the three measurement arms, $\eta_\mathrm{i1}$, $\eta_\mathrm{s2}$ and $\eta_\mathrm{i2}$, into account. From separate measurements we inferred the individual PDC conversion efficiencies, $P_\mathrm{PDC,1}=\left(8.1\pm0.1\right)\times10^{-8}$ and $P_\mathrm{PDC,2}=\left(2.7\pm0.1\right)\times10^{-7}$ pairs per pump photon. Additionally, the injection efficiency of the pump into the waveguide structure is around $\eta_\mathrm{p}^\mathrm{in}=0.5\pm0.1$. By calculating
\begin{equation}
\label{eq:04}
P^\mathrm{th}_\mathrm{triplet}=\frac{\eta_\mathrm{i1}\eta_\mathrm{s2}\eta_\mathrm{i2}\cdot P_\mathrm{PDC,1} P_\mathrm{PDC,2}\mathbf{P}_\mathrm{p}\eta_\mathrm{p}^\mathrm{in}\lambda_\mathrm{p}}{h c_\mathrm{vac} R_\mathrm{rep}}
\end{equation}
with the vacuum speed of light $c_\mathrm{vac}$, we get a theoretical success probability of $P^\mathrm{th}_\mathrm{triplet}=\left(6.35\pm1.5\right)\times10^{-11}$ per pulse. This is in excellent agreement with the experimentally derived benchmark. Note that the scalability of our source is inherent to Eq. (\ref{eq:04}), because for identical pulse energies the ratio $\mathbf{P}_\mathrm{p}/R_\mathrm{rep}$ and $P^\mathrm{th}_\mathrm{triplet}$ do not change, whereas the absolute number of successfully detected photon triplets increases with higher repetition rates of the pump laser. The pulsed excitation in general and in conjunction with the emission wavelengths of the triplet states make our source fully compatible with existing synchronized telecom infrastructure.

In comparison to other approaches on tripartite-state generation \cite{Hamel2014,Ding2015,Khoshnegar2015}, our integrated source offers around two to four orders of magnitude less photon-triplets per unit time. The main limiting factors on the detected photon triplets rates in our setup are given by the necessity of spectral filtering in the secondary PDC output and by the individual conversion efficiencies of the two PDC stages. Further improvements of our waveguide technology can reduce their impact on the success probability in the future. We expect an increase of detected photon-triplet rates of at least one order of magnitude solely by accessing the full spectral bandwidth of the secondary PDC outcome. The resulting, spectrally multimode, secondary photon pairs can be deployed for example for absolute calibration of broad-band-sensitive single photon detectors in the telecommunication bands. Besides, the strong temporal correlations of our spectrally multi-mode secondary PDC photons offer energy-time-entanglement. This could be combined in future work with time-bin-entanglement schemes and would allow for the heralded generation of hyper-entangled Bell-states.

The implementation of reverse proton-exchanged waveguide structures \cite{Parameswaran2002} could also increase the PDC conversion efficiencies, each by at least one additional order of magnitude. Finally, higher repetition rates of the pump laser, e. g. by temporal multiplexing \cite{Broome2011} will lead to increasing numbers of detectable photon triplets and indicate the scalability of our integrated device. Note that state-of-the-art detector recovery times of around $75\,\mathrm{ns}$ for highly efficient MoSi-based SNSPDs \cite{Verma2015}, in conjunction with detection efficiencies of $\eta_\mathrm{det}\sim87\%$, are still the limiting factor to date rather than available repetition rates for the pump laser. By contrast, increasing the pump pulse energy will not improve the output of photon triplets at high CAR-values, because of the growing impact of higher-order photon contributions. Assuming an identical performance of our source in combination with three high-efficiency MoSi detectors, we are actually limited to pump repetition rates of around $62\,\mathrm{MHz}$ without losing photons due to detector recovery effects. The corresponding gain in terms of photon triplet detection rate would be about 6.2 as compared to this work. Another factor of 6 can be achieved due to the increased efficiency of those novel detectors. Summarizing these sources for improvement we expect to increase the photon-triplet verification rates by around four to five orders of magnitude in future work. This will let the application of pulsed and integrated cascaded parametric down-conversion sources in the field of quantum communication get into reach. Note also that our device does not yet represent a source for multi-partite entanglement. Slight technical variations can be made in order to generate polarization-entangled GHZ-states on-chip, such as replacing our type-0 PDC sources with cascaded type-II PDC stages, each having interlaced poling structures \cite{Herrmann2013}. Additional guided-wave polarizing beam splitters and electro-optical polarization controllers could support these integrated devices.

\section{Conclusion}

In conclusion, our monolithic photon-triplet source demonstrates the strengths of integrated quantum optics in second-order nonlinear materials in terms of robustness to environmental influences and state preparation with high signal-to-noise ratios and coincidences-to-accidentals-ratios. Our integrated device marks important progress towards scalable, miniaturized and reconfigurable quantum circuits with high integration densities, long-term stability and the mutual compatibility with the infrastructure of existing and future quantum networks.

The fundamental dependence of the cascaded triplet generation on higher-order photon contributions also offers new ways for studying decoherence at the transition between the micro- and the macro-world. Seeding our primary PDC process with synchronized weak coherent light at idler 1 wavelengths, for example, provides the generation of single-photon-added coherent states \cite{Agarwal1991} paired with two single photons. This lies at the heart of micro-macro-entanglement and allows an integrated approach, e.g. for the generation of Schroedinger-cat-like states \cite{Schroedinger1935}. Likewise, the monolithic generation of heralded exotic quantum states, in conjunction with the opportunity to add fast optical switches to the very same chip, paves the way for future prospects of quantum communication and quantum network technology.\\

\section*{Acknowledgments}

The authors thank R. Ricken, H. Suche, K. Shalm and T. J. Bartley for fruitful discussions and A. J. Miller, V. Ansari and G. Harder for their help with the detection apparatus. This work has been supported by the German Research Foundation (DFG) (Graduiertenkolleg 1464 "Mikro- und Nanostrukturen in Optoelektronik und Photonik").\\

\section*{Appendix A: Testing the mutual compatibility of two parametric down-conversion processes on-chip}

Parametric down-conversion (PDC) processes require energy conservation of pump (p), signal (s) and idler (i) photons. This can be expressed in the frequency notation
\begin{equation}
\label{eq:app01}
\hbar\omega_\mathrm{p}=\hbar\omega_\mathrm{s}+\hbar\omega_\mathrm{i},
\end{equation}
and in the wavelength notation
\begin{equation}
\label{eq:app02}
\frac{1}{\lambda_\mathrm{p}}=\frac{1}{\lambda_\mathrm{s}}+\frac{1}{\lambda_\mathrm{i}}.
\end{equation}
Likewise, the conservation of momentum must be fulfilled, which is commonly referred to as phase-matching and expressed by
\begin{equation}
\label{eq:app03}
\Delta\vec{k}=\vec{k}_\mathrm{p}-\vec{k}_\mathrm{s}-\vec{k}_\mathrm{i}.
\end{equation}
In waveguides, which are usually dispersive, the co-linear propagation reduces the vectorial notation to a wave-number representation, $\vec{k}_i\rightarrow k_i=2\pi n_{\mathrm{eff},i}/\lambda_i$. The effective refractive indices of the guided waves are given by $n_{\mathrm{eff},i}$. Guided-wave PDC requires the compensation of a phase-mismatch $\Delta k$ between the three involved photons at wavelengths $\lambda_\mathrm{p}$, $\lambda_\mathrm{s}$ and $\lambda_\mathrm{i}$. By periodically inverting the nonlinear susceptibility with period $\Lambda_\mathrm{G}$, we create a rectangular grating with the corresponding wave-number $\Delta k=2\pi/\Lambda_\mathrm{G}$. We include this to the momentum conservation condition:
\begin{equation}
\label{eq:app04}
k_\mathrm{p}-k_\mathrm{s}-k_\mathrm{i}\pm\frac{2\pi}{\Lambda_\mathrm{G}}=\Delta k=0.
\end{equation}
This expression facilitates quasi-phasematching of almost arbitrary wavelength combinations. The effective refractive indices are typically dependent on the temperature of the waveguide material. Thus, the PDC emission wavelengths can be tuned by thermal manipulation.

In our integrated lithium niobate chip, we aim for TM$_{00}$ mode-conversion of green picosecond pump photons (p) to a pair of TM$_{00}$ photons at around $\lambda_\mathrm{s1}=790.5\,\mathrm{nm}$ (signal 1) and $\lambda_\mathrm{i1}=1625\,\mathrm{nm}$ (idler 1). The signal 1 photon can subsequently decay to TM$_{00}$ ``granddaughter'' photon pairs, signal 2 and idler 2, with a spectral distribution of about $\Delta\lambda\approx\pm35\,\mathrm{nm}$ around the degeneracy wavelength $\lambda_\mathrm{s2/i2}=1581\,\mathrm{nm}$. The overall cascaded PDC process is described by the formula
\begin{equation}
\label{eq:app06}
\begin{array}{ccccccc}
\mathrm{p}&\rightarrow&\mathrm{i1}&+&\mathrm{s2}&+&\mathrm{i2},\\
532\,\mathrm{nm}&\rightarrow&1625\,\mathrm{nm}&+&(1581\mp35)\,\mathrm{nm}&+&\left(1581\pm35\right)\,\mathrm{nm}.
\end{array}
\end{equation}

Before setting up our cascaded PDC process, the individual PDC sections have been characterized thoroughly, because the intermediate signal photons will serve as the pump for the secondary PDC process. This means that we have to make both processes mutually compatible, since the chosen poling periods are fixed and allow only for raw setting of the quasi-phase-matching conditions. Thus, we acquired the spectra of the signal photons of the primary PDC process at different temperatures using a commercial fiber-coupled spectrometer system and deduced the temperature tuning curve of the first PDC process. 

Additionally, we characterize the secondary PDC stages in two different ways. First, we make use of the fact that second harmonic generation (SHG) represents the reverse three-wave mixing process of degenerate PDC. Therefore, we inject coherent fundamental light from a tunable external cavity laser at wavelengths $1570\,\mathrm{nm}\le\lambda_\mathrm{F}\le1610\,\mathrm{nm}$ to our waveguide structures and measure the SHG with a photo-diode. In order to fine-tune the secondary quasi-phasematching condition, we also focus on the temperature-dependent behavior of the SHG peak wavelength.

\begin{figure}[t]
\includegraphics[width=0.48\textwidth]{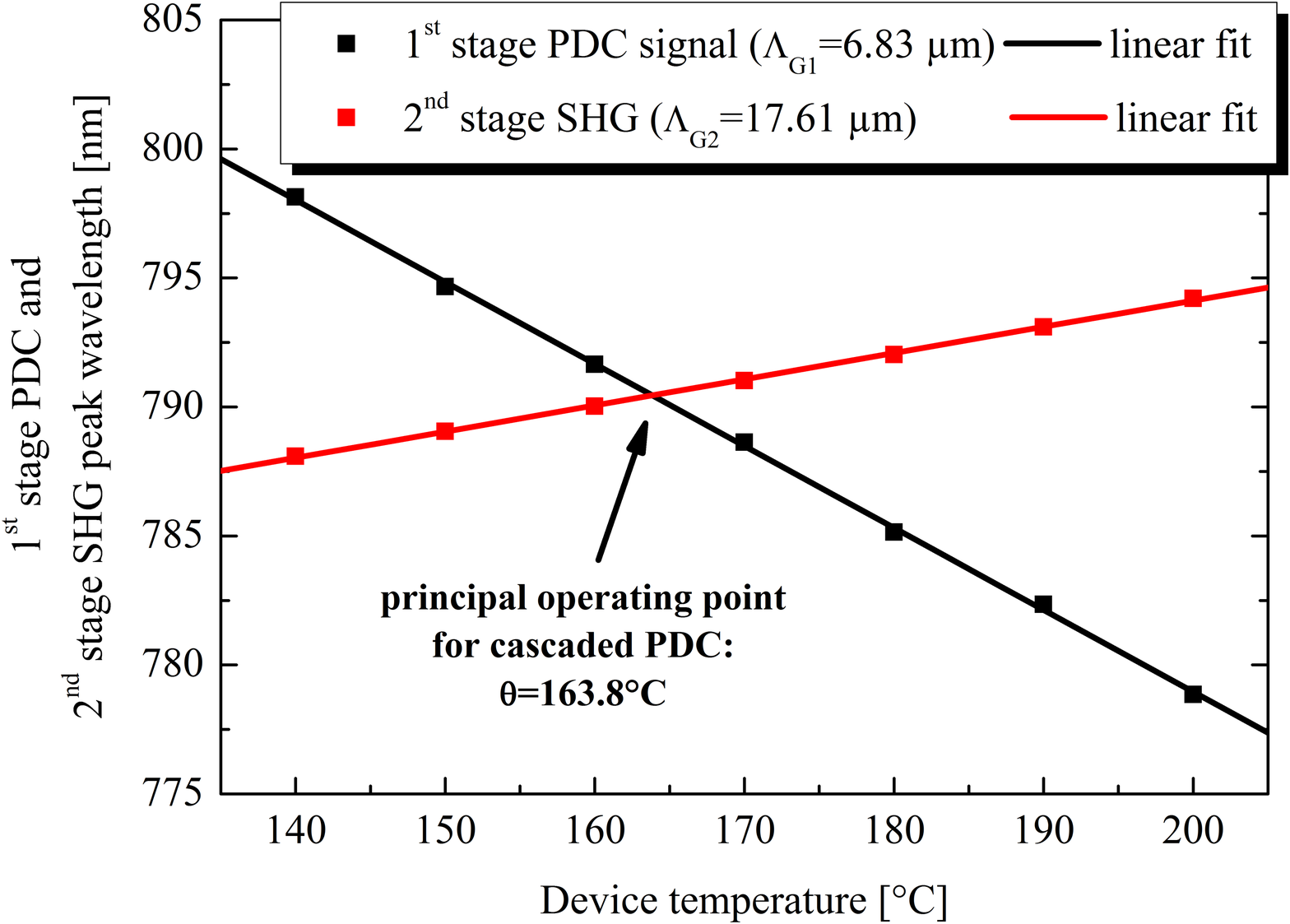}
\includegraphics[width=0.5\textwidth]{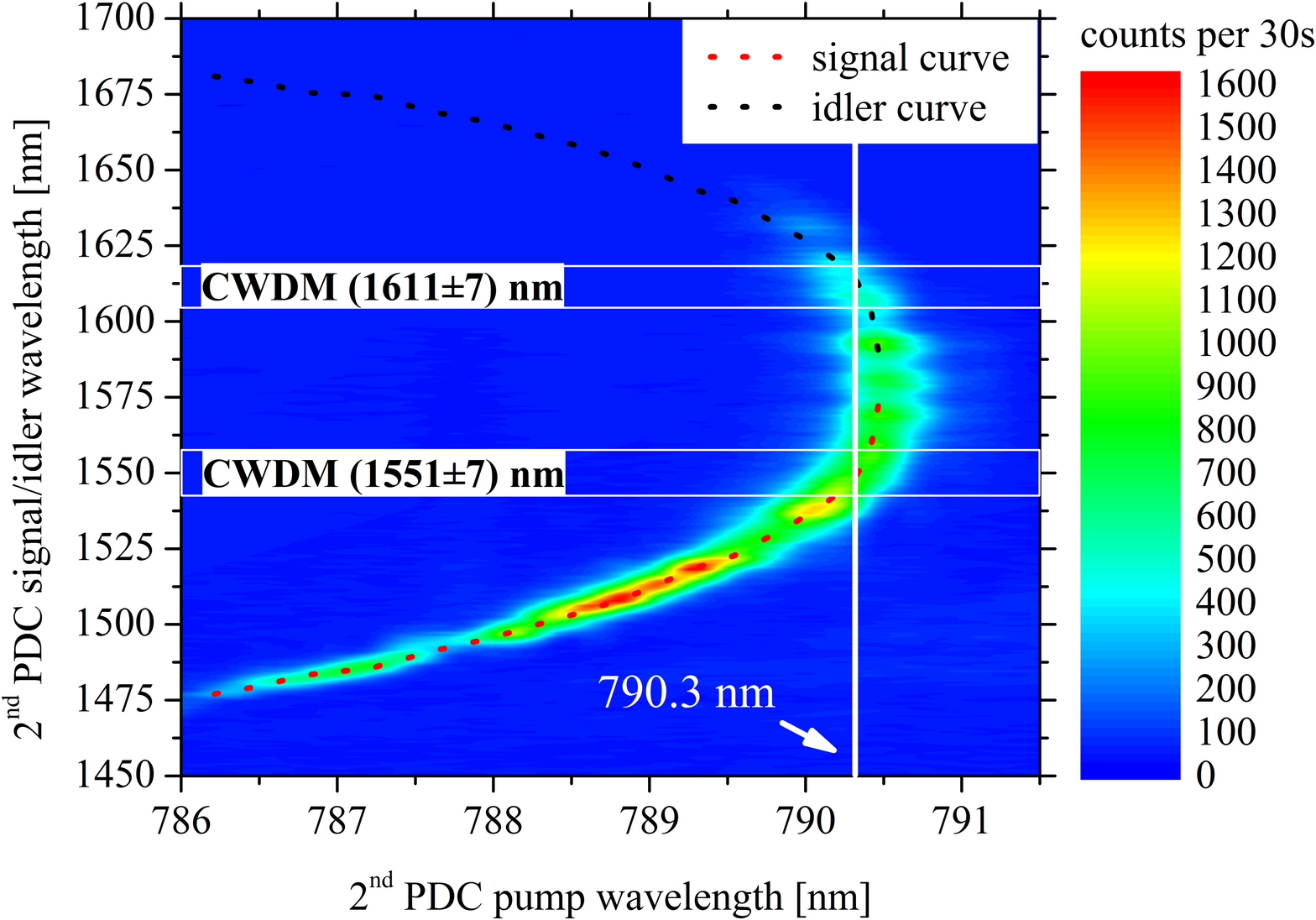}
\caption{\label{SHG-PDC-tuning-temp}Determination of the optimum operation point for verifying photon triplets of high purity: with the two individual temperature tuning curves for fixed poling periods, it is convenient to find the fundamental point of operation at $\theta=163.8^\circ\mathrm{C}$ and $\lambda_\mathrm{s1}=790.5\,\mathrm{nm}$ (left). The spectral splitting of signal and idler wavelengths can be tuned such that addressing of suitable fiber-based filters (CWDM) can be achieved (right). This operation condition is fulfilled, when the secondary PDC is pumped with signal 1 photons at $\lambda_\mathrm{s1}=\lambda_\mathrm{p2}\le790.3\,\mathrm{nm}$. Note that the idler photons tend to be guided more and more weakly at wavelengths above $1635\,\mathrm{nm}$ due to the cut-off-condition of our waveguides.}
\end{figure}

With the temperature tuning curves of both down-conversion stages stages at hand, we extract a \textit{principle} operation temperature (POT) for the cascaded parametric down-conversion process. At $\theta^\mathrm{POT}=163.5^\circ\mathrm{C}$, we observe a signal 1 wavelength of $\lambda_\mathrm{s1}=\lambda_\mathrm{p2}=\left(790.47\pm0.35\right)\,\mathrm{nm}$, which is shown in Fig. \ref{SHG-PDC-tuning-temp} (left). However, at this temperature-wavelength-combination the secondary PDC emission will be degenerate with broad spectral distribution. But for our photon-triplet detection we want to split the secondary photons quasi-deterministically. A non-degenerate operation is beneficial to achieve this. As the second characterization method we therefore performed the direct generation of secondary PDC photon pairs at the fixed device temperature $\theta=\left(163.5\pm0.1\right)^\circ\mathrm{C}$. Picosecond pulsed laser light is deployed in the range of $785.84\,\mathrm{nm}\le\lambda_\mathrm{p2}\le791.88\,\mathrm{nm}$ and in steps of $\Delta\lambda_\mathrm{p2}=0.23\,\mathrm{nm}$. We use a highly dispersive fiber in conjunction with a superconducting nanowire single photon detector (Opus One, Quantum Opus) and a time-tagging module (TTM8000, Austrian Institute of Technology) in order to build a calibrated spectrometer\cite{Avenhaus2009a}. This system stretches the unfiltered signal 2/idler 2 pulses in time, according to their spectral components\cite{Valencia2002,Brida2006,Baek2008}.

\begin{figure}[t]
\centering
\includegraphics[width=0.75\textwidth]{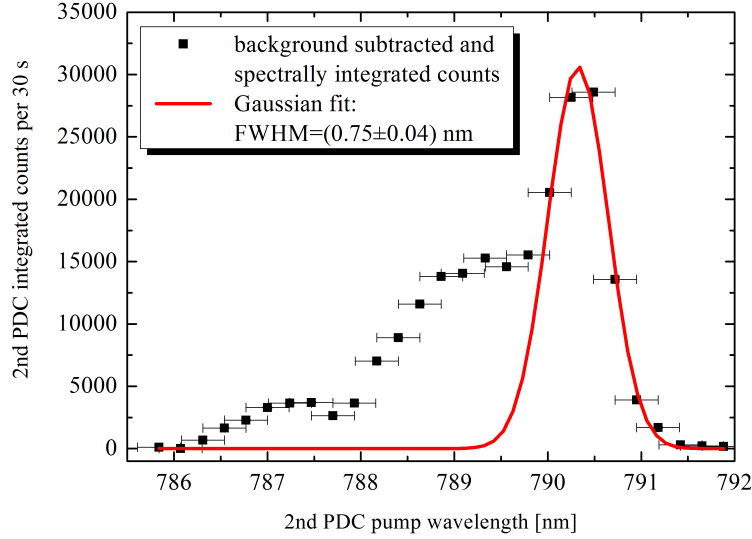}
\caption{\label{PDC-acceptance} Estimation of the acceptance bandwidths of the $2^\mathrm{nd}$ PDC process. We integrate the spectrally resolved PDC outcome dependent on the pump wavelength. The peak at $\lambda_\mathrm{p2}=790.5\,\mathrm{nm}$ denotes the principle wavelength operation point for a fixed temperature, while the width of our Gaussian fit curve indicates the acceptance bandwidth of the secondary PDC process and, thus, defines the good spectral overlap to the first PDC process.}
\end{figure}

We acquire the numbers of click events from the secondary PDC source for $30$ seconds and plot the outcomes pump-wavelength-dependent and color-coded in Fig. \ref{SHG-PDC-tuning-temp} (right). When pumping at $\lambda_\mathrm{p2}=\left(790.49\pm0.23\right)\,\mathrm{nm}$, we identify signal 2 and idler 2 at degenerate wavelengths of $\lambda_\mathrm{s2}=\lambda_\mathrm{i2}=\left(1581.0\pm0.5\right)\,\mathrm{nm}$, as we expected it. We also infer from the graph that, with decreasing pump wavelengths, the PDC emission splits into two arms of non-degenerate signal and idler wavelengths. The spectral bandwidth of the signal arm narrows down at shorter pump wavelengths. The same holds true for idler photons due to energy conservation. The graph does not provide this feature, because the idler photons tend to be weakly guided at wavelengths higher than $\lambda_\mathrm{i2}\ge1635\,\mathrm{nm}$. This effect could be reduced by dispersion engineering of our waveguides.

In order to prevent idler photon scattering to the lithium niobate substrate, we inferred $\lambda_\mathrm{s1}=\lambda_\mathrm{p2}=790.3\,\mathrm{nm}$ as the optimum pump wavelength for the secondary PDC process. This has also the advantage, that signal and idler emission are concentrated in the wavelength regions around $\lambda_\mathrm{s2}=\left(1551\pm25\right)\,\mathrm{nm}$ and $\lambda_\mathrm{i2}=\left(1611\pm25\right)\,\mathrm{nm}$, respectively. That choice allows us to use fiber-based coarse wavelength division multiplexers (CWDM) with very good filtering properties for unwanted wavelengths. Note, however, that these filterws have a narrower transmission bandwidth than our PDC emission, and  we will reduce the detectable event rates in turn.

Additionally, we estimate the pump acceptance bandwidth of our secondary PDC stage, which should spectrally match with the primary signal wavelength in the cascaded process. Integrating the individual $2^\mathrm{nd}$-stage PDC emission spectra over time and subtracting the integral noise background results in the graph in Fig. \ref{PDC-acceptance}, where we plot the dependency on the pump wavelength. The accumulated emission shows a maximum at the degeneracy point $\lambda_\mathrm{p2}=790.5\,\mathrm{nm}$. The data points at short pump wavelengths reflect non-degenerate PDC. By contrast, the steep drop above the degeneracy pump wavelength indicates the tendency to non-phase-matched cases.

From the Gaussian fit, we deduce a spectral acceptance bandwidth of $\Delta\lambda^\mathrm{FWHM}_\mathrm{p2}=\left(0.749\pm0.054\right)\,\mathrm{nm}$. This value is narrower than what we measured for the emission of the primary PDC signal photons, which means that the spectral overlap of the two processes was limited to $\eta_{\lambda_\mathrm{s1}-\lambda_\mathrm{p2}}=0.88$.  Further technological improvement will allow for the exact matching of the two bandwidths, e. g. by adapting the effective lengths of the two involved periodically poled areas.

As the final characterization result, we extract the optimum operating temperature of $\theta^\mathrm{opt}=163.8^\circ\mathrm{C}$ for the cascaded PDC process. We take the temperature tuning curves of each PDC stage into account and consider the desired non-degenerate emission of secondary PDC photons for optimized filtering. At this operating temperature, the intermediate primary signal wavelength is stabilized at $\lambda_\mathrm{s1}=\lambda_\mathrm{p2}=\left(790.3\pm0.032\right)\,\mathrm{nm}$, which provides very good spectral overlap of the two individual PDC processes.

\section*{Appendix B: Histogram of the absolute frequencies of three-fold coincidences}

In Fig. \ref{statistics} of the main text, we fit the absolute frequencies of the detected three-fold coincidences per time bin in terms of a Poisson distribution. Because our data analysis software does not provide factorials, we calculate them by using the gamma function $N!=\Gamma(N+1),\;n\in\mathbb{N}$. The Poisson fit method is reasonable here, but requires additional explanation at this point.

We are aware that each detector might show an individual (probably thermal) dark count statistics. Additional contributions to our noise floor, ~i. e. the emitted blackbody radiation of our source, is also expected to be distributed with thermal statistics. However we expect that our actual three-fold coincidence statistics involve several modes and detectors such that their convolution results in a Poisson distribution. Moreover, in the limit of small expectation values or mean values of the absolute frequency of three-fold coincidence per time bin, both distributions tend to behave identical. Thus we consider the description of our rates by means of Poisson statistics to be adequate. The three-fold coincidence rate for noise and accidentals is hence modeled by the convolution of three Poisson distributions. By contrast, the generation of the overall photon triplets is also assumed to follow Poisson distribution, since they stem from cascaded multi-mode PDC processes. But the triplets manifest themselves by strong correlations for the three-fold coincidence rates, because three photons will be generated for every cascaded PDC event. We note that it is worth to model our system in terms of the exact statistical behavior. But since the description of the individual detector responses is typically non-trivial, it should be given elsewhere in more detail.

\end{document}